# White paper of the "soft X-ray imaging spectroscopy"

Proposers: Noriyuki Narukage (NAOJ; noriyuki.narukage@nao.ac.jp), Shin-nosuke Ishikawa (ISAS/JAXA), Tomoko Kawate (ISAS/JAXA), Shinsuke Imada (Nagoya University), and Taro Sakao (ISAS/JAXA)

<u>Science objectives</u>    The solar corona is full of dynamic phenomena, e.g., solar flares, micro flares in active regions, jets in coronal holes and in the polar regions, X-ray bright points in quiet regions, etc. They are accompanied by interesting physical processes, namely, magnetic reconnection, particle acceleration, shocks, waves, flows, evaporation, heating, cooling, and so on. The understandings of these phenomena and processes have been progressing step-by-step with the evolution of the observation technology in EUV and X-rays from the space. But, there are fundamental questions remain unanswered, or haven't even addressed so far. For example,

- ✓ For particle acceleration,
    - ➢ How are the particles accelerated in the solar flare in the reconnecting magnetic field structure? What is the spatial and temporal relationship between the site of particle acceleration and the MHD structures in the flaring magnetic fields such as shocks and the reconnection region?
        - ✧ The acceleration site has not yet been clearly identified (Ref. 1-7).
- ✓ For further understanding of the magnetic reconnection,
    - ➢ Are there any shocks in the reconnecting magnetic structure of flares and coronal jets?
        - ✧ Though the presence of the shocks is expected (Ref. 8), there is no critical observational evidence of the shocks (Ref. 9-10).
    - ➢ (If the shock exists,) What are the physical quantities of the shock?
    - ➢ How much energy is spent for each physical process of the magnetic reconnection, including particle acceleration?
- ✓ For the understanding of the coronal heating,
    - ➢ Are there > 10 MK components in the non-flaring active region?
        - ✧ Though there are some studies trying to put the upper limit on high temperature components (Ref. 11-12), extensive survey of high temperature components is yet to be conducted.
    - ➢ Are there high energy components ubiquitously present even in the quiet sun?
    - ➢ (If high energy components exist,) Is there any particular region where the heating is concentrated? If yes, what is the photospheric/chromospheric counterpart (e.g., spicule, wave, magnetic cancellation, and so on; Ref. 13-14)?
    - ➢ (If high energy components exist,) Is it enough for the steady coronal heating up to 1 MK?

Our scientific objective is to understand underlying physics of dynamic phenomena in the solar corona, covering some of the long-standing questions in solar physics such as particle acceleration in flares and coronal heating.

<u>Observation method</u>    In order to achieve these science objectives, we need to investigate the processes of the energy storage and its dissipation in the corona in detail. For this investigation, measurement of the physical quantities, especially, temperature and density, with suitable temporal and spatial resolutions is required. In particular, the capability of the detection of higher energy components than the ambient plasmas is indispensable. We give concrete and detailed explanations for it. Since it is expected that most of energetic events (except for flares) generate small amount of the high energy plasma in the bulk ambient plasma, high sensitivity to high energy plasma more than several keV (i.e., high temperature plasma beyond 10 MK) is required. Additionally, since it is also expected that such energetic plasma is rapidly generated and rapidly dissipated, the electron temperature from continuum X-ray emission, which reflects the instantaneous temperature, should be measured. (Note: There is a time gap between the exact temperature and the formation of the corresponding lines, since the ionization equilibrium takes a certain time (Ref. 15-17).)

Considering the above, we identify the ***imaging spectroscopy*** (the observations with spatial, temporal and energy resolutions) ***in the soft X-ray range*** (from ~0.5 keV to ~10 keV) is a powerful approach for the detection and analysis of energetic events (Ref. 18). This energy range contains many lines emitted from below 1 MK to beyond 10 MK plasmas plus continuum component that reflects the electron temperature. The advantages of continuum observation in this energy range are as follows:
- ✓ The continuum is significantly useful to detect the rapid change in the temperature.
- ✓ The energy where thermal and non-thermal components are separated in the energy spectrum distribution is expected to be located in the range of less than 10 keV.

These advantages are significantly valid for the detection of the onset of the high temperature and/or high energy components. Furthermore, the lines located in this energy range also have following advantages for the study of high energy plasmas.
- ✓ The lines located around 6.8 keV (Fe XX – Fe XXVI) are sensitive to the plasma above 10 MK. Though these lines cannot be resolved with this instrument, we can identify the contribution of each lines by the fitting of the blended spectrum of these lines.
- ✓ The neutral iron K$\alpha$ emission (at 6.4 keV) emitted from the X-rays of above 7.1 keV is expected to be a good indicator of the non-thermal electrons (Ref. 19).

The combination of these continuum and lines will surely give us rich information about the high energy plasma. Despite wealth of information it can provide, there has been little imaging spectroscopic observation of the soft X-ray corona conducted thus far (Ref. 20). In this regard, we believe there is rich discovery space with our proposed soft X-ray imaging spectroscopy.

<u>Requirements for the instrument</u>    Table 1 is the summary of requirements for this instrument. The top priority is the coronal observation with all of the spatial, temporal and energy resolutions with an energy sensitivity from ~ 0.5 keV to ~10 keV.

[Temporal resolution] In order to detect the energetic events, the temporal resolution should have the highest priority among these three resolutions, since the energetic event rapidly proceeds and plasma situation (physical quantities) quickly and dynamically varies.

Table 1: Requirements for the instrument

| | |
|---|---|
| ***Temporal sampling*** | 1 msec for > (100 arcsec)$^2$ FOV |
| ***Energy range*** | ~ 0.5 keV – ~ 10 keV |
| ***Energy resolution*** | <~ 100 eV @ 1 keV <br> <~ 200 eV @ 5 keV |
| ***Spatial sampling*** | <~ 1 arcsec |

[Energy resolution] Though the energy resolution in Table 1 is determined by the Silicon CMOS detector, which is a key device of this instrument as described later, this resolution is acceptable to detect the energetic events with both continuum and lines (see Figure 1).

[Spatial resolution] The spatial resolutions in Table 2 are examples for the study of the particle acceleration site and the observation of loops, where spatial binning is done for the good signal-to-noise ratio data with high temporal and energy resolutions. While in this table we show the example for a flare and active region, we can also perform the spectroscopic observation for the quiet sun with larger spatial and/or temporal binning. Since the spatial sampling is < 1 arcsec, this instrument can also take high-spatial-resolution coronal X-ray images better than Hinode/XRT with moderate temporal and/or energy resolution.

To observe soft X-rays, a space-borne instrument aboard a satellite is a must, since the soft X-rays cannot reach to the ground nor to the flight altitude of a balloon.

Table 2 – Expected resolutions to make a spectrum with 20 % statistical error (like Figure 1)

| | | |
|---|---|---|
| For particle acceleration site | ***Temporal resolution*** | < several seconds |
| | ***Resolvable spectral component*** | Multi-thermal + non-thermal |
| | ***Spatial resolution*** | <~ 4 arcsec |
| For flaring & active region loop | ***Temporal resolution*** | < several ten seconds |
| | ***Resolvable spectral component*** | Multi-thermal |
| | ***Spatial resolution*** | <~ 2 arcsec |

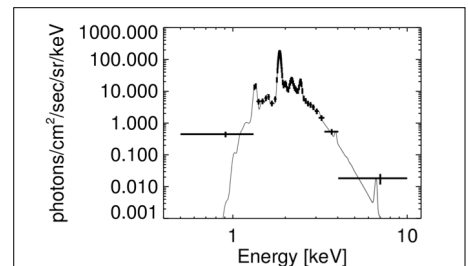

Figure 1 – Expected spectrum of an active region, where the number of energy bin is 80, and the statistical error is 20% in each energy bin.

**Design of the instrument**   This instrument consists of a Wolter Type-I X-ray telescope and a high speed soft X-ray camera (back illuminated CMOS camera) as shown in Figure 2.

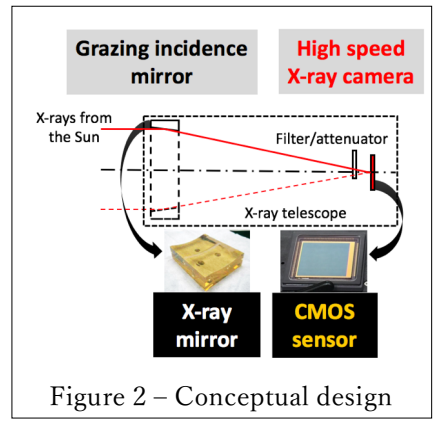

Precision grazing-incidence mirrors with Pt or Ir coating efficiently focus the X-rays up to ~10 keV on the camera with little scattered light.

The soft X-ray imaging spectroscopy is realized with the following method. We take images with a short enough exposure to detect only single X-ray photon in an isolated pixel area with a fine pixel Silicon sensor. So, we can measure the energy of the X-ray photons one by one with spatial and temporal resolutions, since the signal generated by an X-ray photon in a sensor is proportional to the energy of the incident X-ray photon. The energy resolution is ~ 200 eV at 5 keV as shown in

Figure 2 – Conceptual design

Table 1 that is determined by the Fano noise of the Silicon sensor and readout noise. When we use the high speed soft X-ray camera that can perform the continuous exposure with a rate of 1,000 times per second, we can count the photon energy with a rate of several 10 photons / pixel / second. For example, by the combination of this camera and a telescope with a spatial sampling of 1 arcsec, we can make a spectrum shown in Figure 1 every 30 seconds with a spatial resolution of 2 arcsec.

**Feasibility of the instrument**   Our team is now developing mirror and camera to realize this instrument.

[Mirror] A precision engineering Wolter mirror whose focal length is 4 m has successfully achieved the focusing spot size of ~0.2 arcsec FWHM and ~3 arcsec HPD (half-power diameter) at 8 keV (Ref. 21). We are now proceeding towards sub-arcsec HPD with fine mirror polishing. We expect that our mirror can reach the level of the X-ray mirrors aboard the Chandra X-ray observatory.

[Camera] Figure 3 is the evaluation result of the photon counting capability of the back illuminated CMOS sensor with an Fe 55 source. The Mn K$\alpha$ (5.9 keV) and K$\beta$ (6.5 keV) lines are clearly separated. We also evaluated this sensor with the synchrotron beam from 0.8 keV to 4.5 keV. On the basis of these evaluations, we have confirmed that this sensor has a photon counting capability at least from 0.8 keV to 6.5 keV. Now, we are developing a high speed soft X-ray camera with this CMOS sensor.

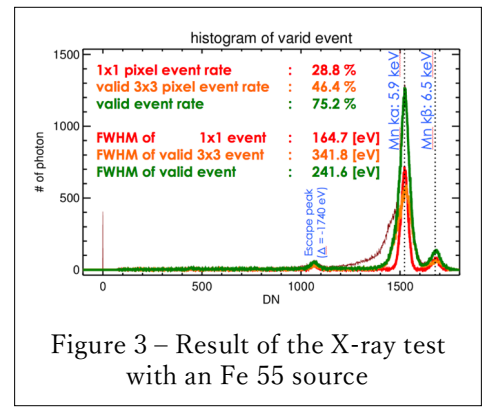

[Sounding rocket experiment] We plan to mount our camera to the FOXSI 3 sounding rocket to perform the solar soft X-ray imaging spectroscopy in the summer of 2018. This is a good opportunity to demonstrate the scientific advantage in the soft X-ray imaging spectroscopy for the coronal study and the feasibility of the instruments for the satellite mission.

Figure 3 – Result of the X-ray test with an Fe 55 source

**Candidates of collaborating instruments**   We consider some candidates of collaborating instruments with our soft X-ray imaging spectroscopic telescope as follows (in the order of targeted energy):
1. Collaboration with "hard X-ray imaging spectroscopy (e.g., FOXSI SMEX)" gives us the continuous energy coverage from 0.5 keV to above 50 keV.
2. Collaboration with "soft X-ray spectroscopy with grating (e.g., MaGIXS sounding rocket)" can compensate for the weakness of each instrument, namely, energy resolution and FOV.
3. Collaboration with "EUV spectroscopy (e.g., EUVST)" provides rich information about the coronal plasma (electron and ion temperatures, non-thermal components, density and velocity).
4. Though the scientific feasibility of the UV spectropolarimetry is under examination, collaboration with "UV spectropolarimetry (e.g., CLASP)" may provide us the relation between the magnetic fields in the solar atmosphere and coronal energetic events.

**Schedule**   We aim to realize the solar soft X-ray imaging spectroscopic observations as a satellite mission at the next solar maximum around 2025.